\documentclass{article}
\usepackage{spconf,amsmath,graphicx,hyperref}

\title{Differentiable Optimization of Linear Differential Microphone Arrays: A
Joint Geometry and Filter Design Framework}
\name{Siminfar Samakoush Galougah, Ramani Duraiswami
}
\address{University of Maryland, College Park}
\begin{document}
%
\maketitle
\begin{abstract}
This paper presents a differentiable optimization framework for the design of constrained Linear Differential Microphone Arrays (LDMAs). The proposed method leverages a non-uniform delay-and-sum beamformer as a light-weight base system model, proving its ability to achieve the optimal beampattern of LDMAs by jointly optimizing microphone positions and filter weights. The formulation enables the optimized design of a filter with a distortion-free constraint in the desired sound direction, while also imposing constraints on microphone positioning to ensure consistent performance. Through evaluation on multiple metrics, including Mean Squared Error (MSE), Directivity Index (DI), White Noise Gain (WNG), and computation time, and comparison with state-of-the-art methods, this approach demonstrates a flexible, directive, robust, and hardware-efficient design.

\end{abstract}
\section{Introduction}
\label{sec:intro}
Differentiable optimization is introduced as a framework in signal processing \cite{wang2021diff}, enabling gradient-based learning of beamforming parameters. Frameworks such as OptNet \cite{amos2017optnet} have demonstrated the feasibility of embedding constrained optimization within differentiable models. Inspired by these works, we propose a novel method for designing LDMAs by formulating a constrained delay-and-sum beamformer using differentiable programming. It jointly learns the filter coefficients and microphone spacings by minimizing a cost function defined over a target beampattern in a fully differentiable pipeline, enabling flexible LDMA design beyond the capabilities of the existing approaches. The optimization process demonstrates that the conventional LDMA beampattern could be achieved by optimally adjusting the delay-and-sum beamformer’s filter weights and microphone positions.

Differential Microphone Array (DMA) often involves a cumbersome multistage process; As the number of microphones increases, the complexity of this process increases significantly. The focus of our work is to explore a lightweight, low-computational-cost architecture with performance comparable to existing methods. Using the delay-and-sum concept and automatic differentiation, we propose a novel design that achieves the desired performance. Our approach accomplishes this by directly optimizing microphone array parameters under specific constraints, resulting in a more efficient and optimized design process, validated through performance metrics such as MSE, DI, and WNG.

Unlike previous approaches, which are insufficient for obtaining filter weights for LDMAs of any order, we present a novel and previously unexplored framework that enables LDMA design of any order by directly optimizing the cost function under specific constraints: (1) ensuring distortion-free signal reception and (2) limiting the distance between microphones. We validated the performance of the proposed approach for designing LDMAs of various orders by comparing it to the optimal desired performance using several metrics, including beampattern, MSE, DI, WNG, and computation time. 
Through evaluation on multiple metrics, this approach demonstrates a flexible, directive, robust, and hardware-efficient design.


\section{Related Work}
Several studies have laid the foundation for the design of LDMAs. For example, \cite{elko1997dma} proposed the classical design of the LDMA with fixed microphone spacing and filter weights, approximating the spatial derivative by the finite difference method. Since this approach lacked flexibility and robustness, authors in \cite{b29} proposed a robust design of LDMAs, which addressed the white noise amplification in wideband signals, improving performance with frequency-invariant beampatterns via Maclaurin’s series expansion. Building on this, \cite{b23} introduces a two-stage robust DMA beamforming approach to maximize WNG, addresses extra-null issues at high frequencies, and offers frequency-independent beampatterns. Similarly, considering key performance metrics such as Directivity Factor (DF) and WNG, \cite{b30} optimized the geometry of LDMA by dividing the frequency band into subbands and using particle swarm optimization, and
expanding on this, \cite{b27} explored nonuniform LDMAs using spatial difference operators and a two-stage design process. Considering steering limitations, \cite{b25} presents differential beamformers with linear microphone arrays, proposing new strategies for designing steerable beamformers. Similarly, \cite{b28} proposed a fully steerable design using both omnidirectional and bidirectional microphones, validated through simulations. To cancel the interference and signals coming from an unwanted direction, \cite{jin2019null} proposed the null-constraint LDMAs by 
introducing a directional constraint to remove the unwanted signals. More recently, \cite{jin2022geometry} considered the geometry optimization of the microphone array with fixed filter weights. As mentioned, none of the previous papers have worked on the joint optimization of geometry and filter weights to achieve an optimal, flexible, and robust design for LDMAs.

\section{System model and Problem Definition}
\label{sec:sphd}
\subsection{System Model of the Nonuniform Linear Microphone Arrays}
\begin{figure}[h]
    \centering
\includegraphics[width=0.45\textwidth]{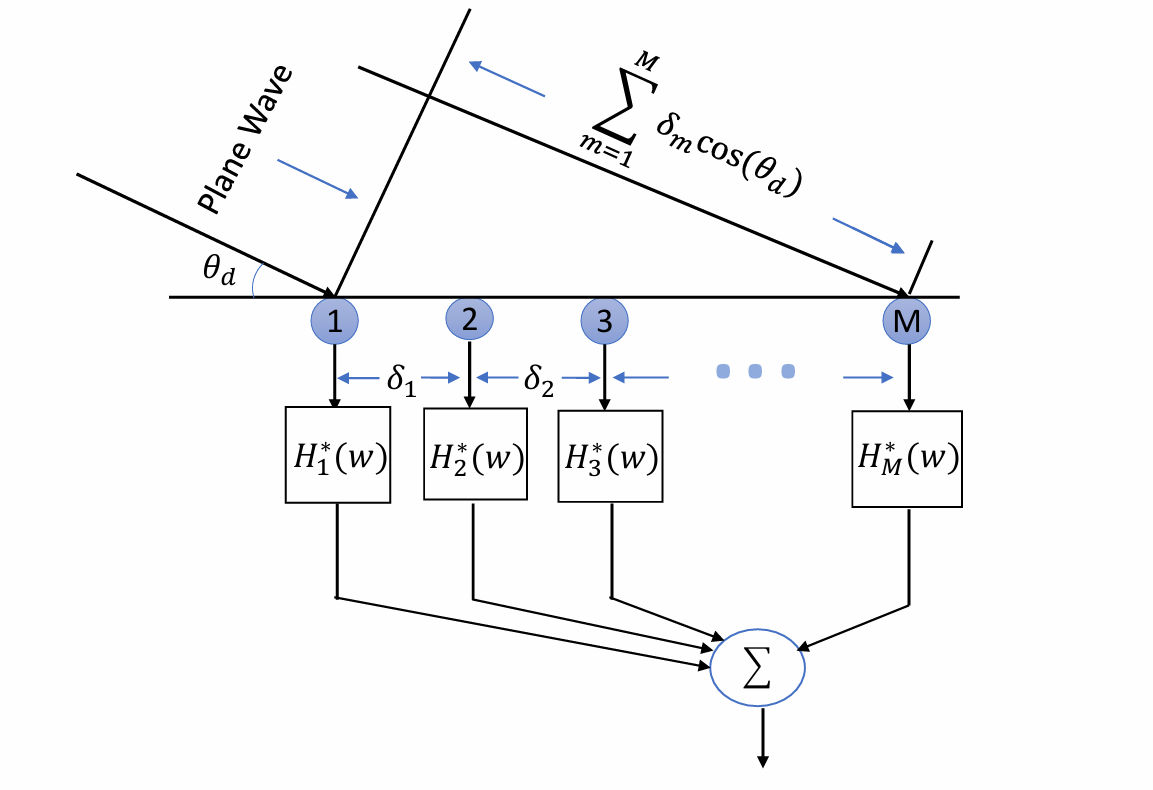}
\caption{A non-uniform linear microphone array.}
    \label{fig:ldma}
\end{figure}

We consider a plane wave coming from a far-field sound source with the speed of $c=340~ m/s$ encountering an nonuniform delay-and-sum beamformer with $M$ omnidirectional microphones. The distances between microphones are not uniform, with the distance between the $m$th microphone and $m+1$th microphone denoted $\delta_{m}$ for $m=1, ..., M-1$. The position of each microphone can be adjusted to achieve the desired beampattern. 
Assuming the desired sound signal coming from the direction which is defined by the relative azimuthal angle $\theta_{d}$, where the main lobe is oriented towards that angle. In this context, the steering vector is as follows   
\begin{equation}
\label{dh}
\begin{split}
\mathbf{d}(w,\theta_{d})=[1~e^{-jw\tau_{1}\cos{\theta_{d}}}~e^{-jw\sum_{i=1}^{2}\tau_{i}\cos{\theta_{d}}}~...\\~e^{-jw\sum_{i=1}^{M-1}\tau_{i}\cos{\theta_{d}}}]^{T}  \end{split} 
\end{equation}
where the $\text{superscript}^{T}$ denotes the transpose operator and
$j$ represents the imaginary unit., $w=2\pi f$ is the angular frequency, and $\tau_{i}=\delta_{i}/c$ for $i=1,...,M-1$ is the delay between microphone $i$ and microphone $i+1$. Fig. \ref{fig:ldma} shows a non-uniform linear differentiable microphone array. As it can be seen, the delay between $m$th microphone and the first microphone is $ \sum_{i=1}^{m-1}\tau_{i}
$ which is the sum of the delays from the first microphone to the $m$th microphone and the respective phase difference would be $\sum_{i=1}^{m-1}w\frac{\delta_{i}}{c}\cos{\theta_{d}}=\sum_{i=1}^{m-1}w\tau_{i}\cos{\theta_{d}}$. 
\subsection{Beampatterns}
We adopt a filter design where the filter coefficients $H_{m}^{*}(w)$ (Note that the superscript 
\(*\) represents the complex conjugate), inter-element spacing $\delta_{m}$, are programmatically determined. 
Our approach optimizes these two variables in the proposed system model using automatic differentiation, ensuring that the resulting beampattern aligns with that of an LDMA of the same order

\noindent The weight $H_{m}^{*}(w)$ is applied to the output of the microphone $m$; the complete set of these weights is represented in the form of a vector containing all individual weights, thereby defining the overall response of the microphone array to incoming sound waves in a vector as follows \begin{equation}
\mathbf{h}(w)=[H_{1}(w)~H_{2}(w)~...~H_{M}(w)]^{T} 
\end{equation}

\noindent Beampattern or directional sensitivity, which represents the sensitivity of microphone arrays to the direction of the incoming sound, is an essential aspect of the array signal processing. It indicates how well the microphone array can capture or reject sound from various directions. The Beampattern of the proposed system model is mathematically defined as follows
\begin{equation}
\begin{split}
&\mathcal{B}_{M}[h(w),\theta]=\mathbf{d}^{H}(w,\theta)\mathbf{h}(w)\\
&=\sum_{m=1}^{M}H_{m}(w)e^{\sum_{i=1}^{m-1}w\tau_{i}\cos{\theta}}
\end{split}
\end{equation}
where $\mathbf{d}^{H}(w,\theta)$ is the steering vector and defined in Eq.
\ref{dh}. 

\noindent For a fair comparison with the LDMA beampattern, the desired beampattern of the Nth-order LDMA, known for its frequency-independent characteristics, is defined as follows:

\begin{equation}
 \mathcal{B}_{d,N}[\theta]=\sum_{n=0}^{N}a_{n}\cos^{n}{(\theta-\theta_{d})},\quad \text{where}~~ \sum_{n=0}^{N}a_{n}=1.  
\end{equation}
where $a_{n}$ for $n=0,...,N$ are real numbers and  $\theta_d$ is the steering direction of the LDMA.
 
\subsection{Optimization Problem}
The objective is to find the optimum weights and positions of the microphones in our beamforming system to minimize the mean squared error (MSE) between the beampattern of the proposed system model and LDMA. This optimization is subject to two constraints: 1) ensuring that the distance $\delta_m$ is within the range of $\delta_{\text{min}}$ and $\delta_{\text{max}}$, and 2) ensuring that the desired sound signal from the direction $\theta_d$ is received without distortion. 
The MSE optimization problem is formulated as follows:
\begin{equation}
\begin{aligned}
&\min_{\delta_{m},H_{m}(w)} ~~\mathrm{E}_{\theta}\big[\big|\mathcal{B}_{d,N}[h(w),\theta]-\mathcal{B}_{M}[h(w),\theta]\big|^{2}\big] \\   
&s.t.~~\mathbf{d}(w,\theta_{d})^{H}\mathbf{h}(w)=1  \\ 
&\quad\quad\delta_{min}\leq \delta_{m}\leq \delta_{max}
\end{aligned} 
\end{equation}
where 
$d(w,\theta_{d})$ is a steering vector (the source signal comes from the direction $\theta_{d}$) defined in Eq.\ref{dh}, then the distortionless constraint at $\theta_{d}$ is as
\[\mathbf{d}(w,\theta_{d})^{H}\mathbf{h}(w)=\sum_{m=1}^{M}H_{m}(w)e^{\sum_{i=1}^{m-1}w\tau_{i}\cos{\theta_{d}}}=1.\]\\
In our computational model, we have the flexibility to set $\delta_{\text{min}} = 0$, allowing for enhanced adaptability in our system's configuration.

\noindent Considering the DF as a way to evaluate the performance of the proposed method, showing how well a Microphone Array directs a sound signal to a specific direction, it is computed as follows 
\begin{equation}
\text{DF}=\frac{|\mathcal{B}_{M}[h(w),\theta_{d}]|^{2}}{\frac{1}{2\pi}\int_{0}^{2\pi}|\mathcal{B}_{M}[h(w),\theta]|^{2}d\theta}.    
\end{equation}
\noindent Note that DI is simply the DF expressed in DB. Therefore, either metric can be used to report the performance of the microphone arrays.

\noindent The MSE between two beampatterns is another way of evaluating the performance of our system, and it is defined as:
\begin{equation}
\mathcal{\epsilon}_{\mathcal{B}_{M}[h(w),\theta]}=\frac{1}{2\pi} \int_{0}^{2\pi}|\mathcal{B}_{M}[h(w),\theta]-\mathcal{B}_{d,N}[\theta]|^{2}d\theta.   
\end{equation}
It shows how well the beampattern of the proposed method is close to the beampattern of an LDMA. 

\noindent In realistic scenarios with sensor noise, the WNG is a useful metric to quantify the robustness of the beamformer to spatially white noise. It is defined as:
\begin{equation}
\text{WNG} = 10 \log_{10} \left( \frac{|\mathbf{d}^H(w, \theta_{d})\, \mathbf{h}(w)|^2}{\|\mathbf{h}(w)\|^2} \right),
\end{equation}

\section{Numerical Results}
Considering the simulation setup where the order of LDMA is $N = 2$, the number of microphones is 
$M = 5$, sound speed is $c = 340 m/s$, 
$f = 2 \text{kHz}$, $
\omega = 2 \pi f = 2 \pi \times 2000 \approx 12,566$. Considering $\tau_{min}=0.0$~\text{second} and $\tau_{max}=0.0009$~\text{second}, we will have the interelement spacing as $\delta_{\text{min}} = 0.0\text{cm}$ and 
$\delta_{\text{max}} = 30.6\text{cm}$. 
The numerical optimization technique used for this constraint problem is the Sequential Least Squares Programming (SLSQP) algorithm, which is part of the minimization function in scipy.optimize the module of the SciPy library in Python. We emphasize that the novelty of this work lies not in the solver itself, but in the differentiable formulation of the LDMA design problem, which allows end-to-end learning of spatial and spectral parameters in a unified optimization pipeline.

\begin{figure}[h]
    \centering
\includegraphics[width=0.43\textwidth]{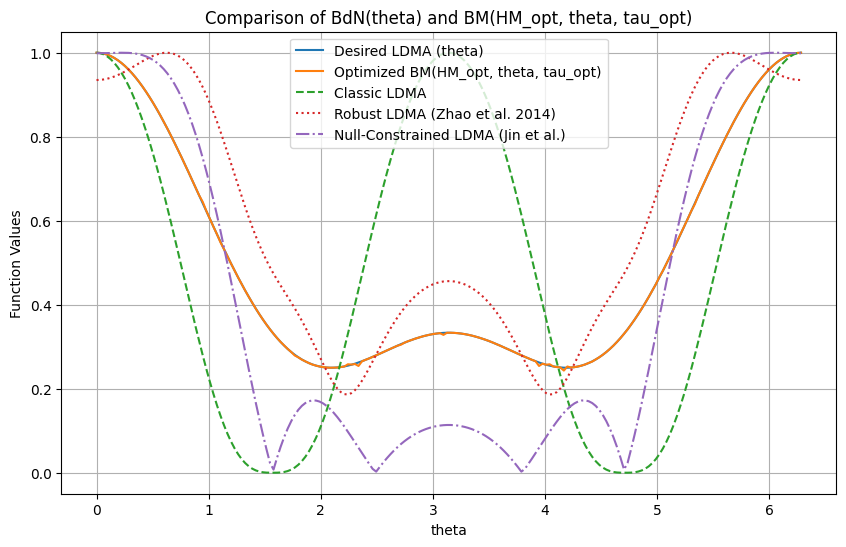}
    \caption{
Comparison of the desired second-order LDMA beampattern 
at
$\theta_{d} = 0$  with three implementations: the proposed optimized method (joint filter and geometry optimization), the classic LDMA design, the robust LDMA, and the null-constraint.
}   \label{fig:td0cmp}
\end{figure}

In Fig. \ref{fig:td0cmp}, we compare our proposed differentiable LDMA design framework for the case where the sound signal from  $\theta_{d}=0$ with the desired LDMA beampattern and the three state-of-the-art approaches: classical, robust, and null-constrained LDMA. The classical LDMA approach \cite{elko1997dma}, despite having low computational cost, has limited flexibility and a considerable mismatch in sidelobe regions with the desired beam pattern. The robust LDMA method \cite{b29} improves the design and reduces white noise amplification. The null-constrained LDMA \cite{jin2019null} enforces directional nulls alongside a unity gain constraint, providing control over interferers. However, these approaches assume fixed geometry and cannot incorporate specific constraints. In contrast, our proposed method is a differentiable constrained optimization problem, jointly learning filter coefficients and microphone positions to match the desired beampatterns. In this figure, you can see that the beampattern produced by the proposed method aligns with the desired beampattern of LDMA over the angular range. Although there are a few small mismatches in some angles, as it is highly dependent on the initial guess of the optimization parameters, $H_{m}(w)$ and $\tau_{m}$.

\begin{table}[h]
\scriptsize
\centering
\caption{Performance comparison of LDMA methods ($N{=}2$, $M{=}5$, $\theta_d{=}0$).}
\begin{tabular}{|l|c|c|c|c|}
\hline
\textbf{Metric} & \textbf{Classic} & \textbf{Robust} & \textbf{Null-Constrained} & \textbf{Proposed} \\
\hline
MSE  & $1.19\mathrm{e}{-01}$ & $2.05\mathrm{e}{-02}$ & $2.88\mathrm{e}{-02}$ & \textbf{$2.54\mathrm{e}{-06}$} \\
DI (dB) & 3.22 & 1.93 & 3.85 & \textbf{2.99} \\
WNG (dB) & $-30.21$ & $-3.70$ & $-8.88$ & \textbf{3.99} \\
Steps & – & – & – & 7 \\
Runtime (s) & – & – & – & \textbf{0.66} \\
\hline
\end{tabular}
\label{tab:ldma_runtime_comparison}
\end{table}

We present the performance of each approach in terms of MSE, DI, WNG, number of steps, and runtime (using CPU) in the Table. \ref{tab:ldma_runtime_comparison}. Experimental results show that our method achieves low MSE around $2.5\times 10^{-6}$, DI close to the desired beampattern $3$ dB, and WNG around $4$dB, which is considered good for a second-order LDMA with five microphones~\cite{b8}. Although there is a trade-off between DI and WNG, both perform well in our method and make a well-balanced design. It is because of the unit-gain constraint in the look direction, which keeps filter weights from becoming excessively large and makes them more robust to white noise. As shown, our method has multiple optimization steps and therefore adds a small runtime overhead. This is not an issue, knowing that in an offline microphone array configuration, processing time is not a bottleneck.

\begin{figure}[h!]
    \centering
\includegraphics[width=0.46\textwidth]{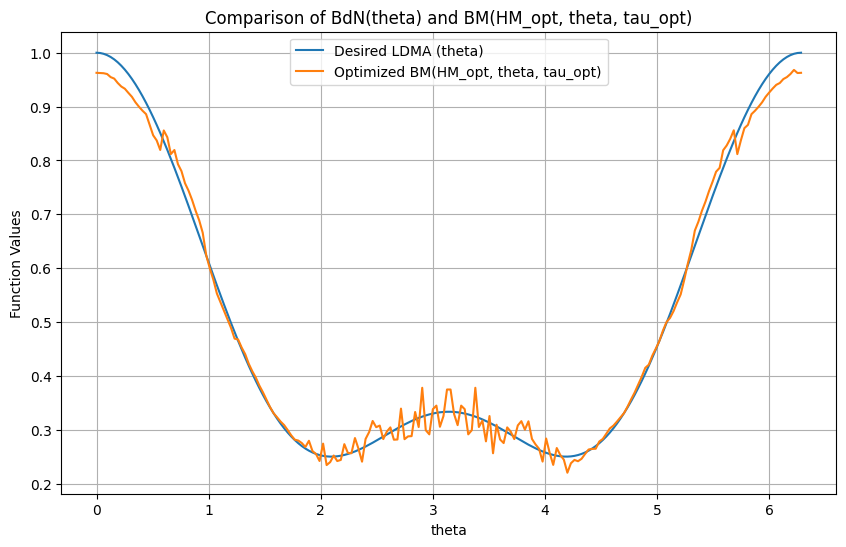}
    \caption{ Desired beam-pattern and the 
 beam-pattern of the proposed method for the second order LDMA when microphones are displaced for $0.1$cm is compared versus the incident angle when the angle of arrival is $\theta_{d} = 0$.}
    \label{fig:rb0}
\end{figure}

The optimized inter-microphone spacings are 0.26, 9.23, 19.87, 30.60 cm. It demonstrates how our method adapts geometry and follows the inter-spacing constraint to shape the desired beampattern. In Fig. \ref{fig:rb0}, we present the beampattern for when the microphones are displaced by $0.1$ cm from the optimal interelement spacing in the proposed methods. The result is noticeably close to the desired beampattern, demonstrating the robustness of the method to small positioning errors.



\begin{figure}[h!]
    \centering
\includegraphics[width=0.42\textwidth]{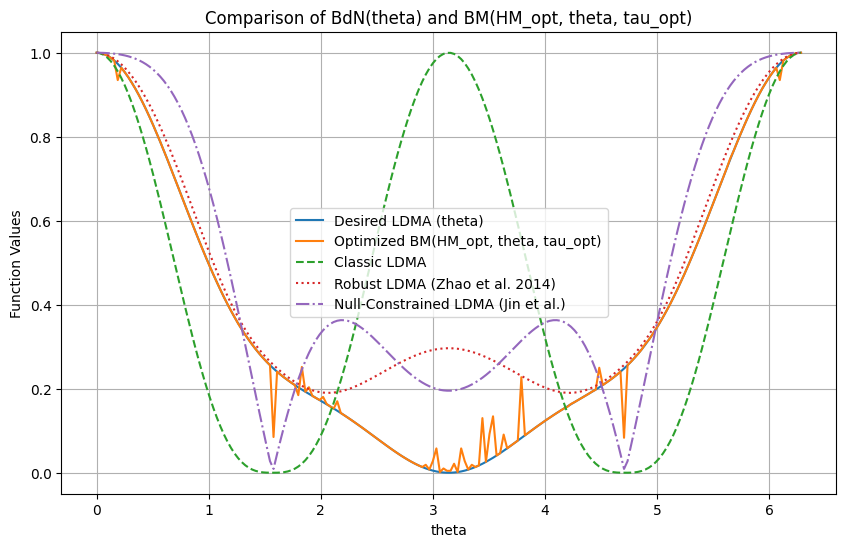}
    \caption{ Comparison of the desired third-order LDMA beampattern 
at
$\theta_{d} = 0$  with three implementations: the proposed optimized method (joint filter and geometry optimization), the classic LDMA design, the robust LDMA, and the null-constraint.}
    \label{fig:2td0}
\end{figure}

Figure~\ref{fig:2td0} evaluates a third-order LDMA $N=3$ at look direction $\theta_d=0$ using the minimal number of microphones, $M=N+1=4$. In Table~\ref{tab:ldma_3rd_runtime_comparison}, we compare the resulting beampattern of the proposed method with the baselines. Even at this minimal array size, the proposed method attains the target response and consistently outperforms the baselines. It indicates that $M=N+1$ is practically sufficient in our proposed design. Therefore, considering implementation cost, we recommend the minimal configuration with $M=N+1$ microphones.

All these results together confirm that with the proposed system model and automatic differentiation as the optimization method, we can achieve the LDMA beampattern of any order with significant performance. 

\begin{table}[!h]
\scriptsize
\centering
\caption{Performance comparison of LDMA methods ($N{=}3$, $M{=}4$, $\theta_d{=}0$).}
\begin{tabular}{|l|c|c|c|c|}
\hline
\textbf{Metric} & \textbf{Classic} & \textbf{Robust} & \textbf{Null-Constrained} & \textbf{Optimized} \\
\hline
MSE         & $1.79\mathrm{e}{-01}$ & $1.38\mathrm{e}{-02}$ & $2.79\mathrm{e}{-02}$ & \textbf{$5.93\mathrm{e}{-04}$} \\
DI (dB)     & 3.48 & 3.46 & 3.10 & \textbf{4.20} \\
WNG (dB)    & $-6.62$ & $-10.07$  & $-8.26$ & \textbf{4.53} \\
Steps       & – & – & – & 6 \\
Runtime (s) & – & – & – & \textbf{0.25} \\
\hline
\end{tabular}
\label{tab:ldma_3rd_runtime_comparison}
\end{table}

Although the current approach focuses on a narrow frequency band, extending the formulation to broadband scenarios is straightforward by incorporating multi-frequency objectives or FIR filters within a differentiable optimization framework.  Extending the framework to broadband signal processing is an important direction for future work.

\section{Conclusion}
This paper proposes a novel approach based on differentiable programming for designing constrained LDMAs of any order, utilizing a lightweight, hardware-efficient delay-and-sum beamformer architecture. The optimization problem is addressed by jointly optimizing two key variables: filter weights and microphone positions, ensuring the resulting beampattern aligns with the desired LDMA beampattern of the same order. Through evaluation on multiple metrics, including Mean Squared Error (MSE), Directivity Index (DI), White Noise Gain (WNG), and computation time, and comparison with state-of-the-art methods, this approach demonstrates a flexible, directive, robust design.

\bibliographystyle{IEEEbib}
\bibliography{strings,refs}

\end{document}